\documentclass[aps,prl,twocolumn,showpacs]{revtex4}
\usepackage{graphicx}
\usepackage{bm}
\usepackage{amsmath}
\begin{document}
\title{Nuclear Spin Relaxation Rate of Disordered $p_x+ip_y$-wave Superconductors}
\author{Qiang Han$^{1,2}$}
\author{Z. D. Wang$^{1,2}$} %
\email{zwang@hkucc.hku.hk}%
\address{$^1$Department of Physics, University of Hong
Kong, Pokfulam Road, Hong Kong, China } %
\address{$^2$Department of Materials Science and Engineering, University
of Science and Technology of China, Hefei 230026,
China}%
\date{\today}

\begin{abstract}

Based on an effective Hamiltonian with
the binary alloy disorder model defined in the triangular
lattice, the impurity scattering effects on the density of states
and especially on the
spin-lattice relaxation rate $1/T_1$ of $p_x+ip_y$-wave
superconductors are studied by solving numerically  the
Bogoliubov-de Gennes equations. In the clean limit, the coherence peak of $1/T_1$
is observed as expected. More intriguingly, for strong scattering potential
, the temperature dependence of $1/T_1$ exhibits the two different
power law behaviors near $T_{\text{c}}$ and at low temperatures,
respectively, which is in good agreement with  the nuclear
quadrupolar resonance measurement.

\end{abstract}
\maketitle

The novel superconductor, Na$_x$CoO$_2$$\cdot y$H$_2$O ($x=0.35$),
recently discovered by Takada {\it et al} \cite{takada} has
stimulated much theoretical and experimental interest in studying
its spin and orbital symmetries of the Cooper pairs, which helps
to explore and understand the underlying superconducting mechanism
of this new material. At present this issue is hotly debated and
there still exist controversies both theoretically and
experimentally. Resonating valence bond \cite{anderson} theories
of the triangular lattice $t$-$J$ model
\cite{baskaran,kumar,qhwang} describe Na$_x$CoO$_2$$\cdot y$H$_2$O
as an electron-doped Mott insulator based on the fact that the
Co$^{4+}$ atoms has spin-$\frac{1}{2}$ as Cu$^{2+}$ in
high-$T_{\text{c}}$ cuprates. Such theories prefer the
spin-singlet $d+id^{\prime}$-wave pairing ($d=d_{x^2-y^2}$ and
$d^\prime=d_{xy}$). On the other hand, theories based on a
combined symmetry analysis with fermiology \cite{tanaka} favor
triplet over singlet pairing with the $\bm{d}$-vector
perpendicular to the cobalt plane [most probably
$\bm{d}=\hat{z}(p_x\pm ip_y)$, i.e. $p_x\pm ip_y$-wave pairing
state]. As for the experimental aspect, a series of nuclear
magnetic/quadruplolar resonance (NMR/NQR) measurements
\cite{kobayashi1,waki,kobayashi2,fujimoto} also give divergent
implication about the pairing symmetry in Na$_x$CoO$_2$$\cdot
y$H$_2$O. Namely, for spin pairing Ref.~\onlinecite{waki} found
the signature for the triplet pairing while
Ref.~\onlinecite{kobayashi2} supported the singlet pairing
although both examined the temperature dependence of the Knight
shift; for the orbital wave function of the Cooper pairs, a
fully-gapped superconducting state \cite{kobayashi1,waki} is
inferred from the existence of the Hebel-Slichter coherence peak
\cite{hebel} of the spin-lattice relaxation rate ($1/T_1$); on the
other hand, opposite conclusion is drawn from the NQR experiment
\cite{fujimoto} that $1/T_1$ decreases across $T_{\text{c}}$
without the coherence peak and follows a power-law dependence
deviating the exponential relation, which implies a nodal gap
function.

Motivated by  the novel power law behavior reported in Ref.\cite{fujimoto} and the fact that
various impurities or defects, such as inhomogeneity in the Na
content, oxygen vacancies and substitution atoms in the CoO$_2$
layer {\it etc.}, are present during the synthesis and handling of
this compound, in this Letter, we  investigate and elucidate  the effect of
disorder on the
nuclear spin relaxation rate of the $p_x\pm ip_y$-wave superconductor
for the first time. Our results
indicate that the seemingly incompatible experimental observations
of the temperature dependence of $1/T_1$ may be reconciled
within the picture of the {\it disordered} chiral $p_x+ip_y$-wave
pairing state if the effect of impurities is properly considered.

\begin{figure}
\includegraphics[width=8 cm]{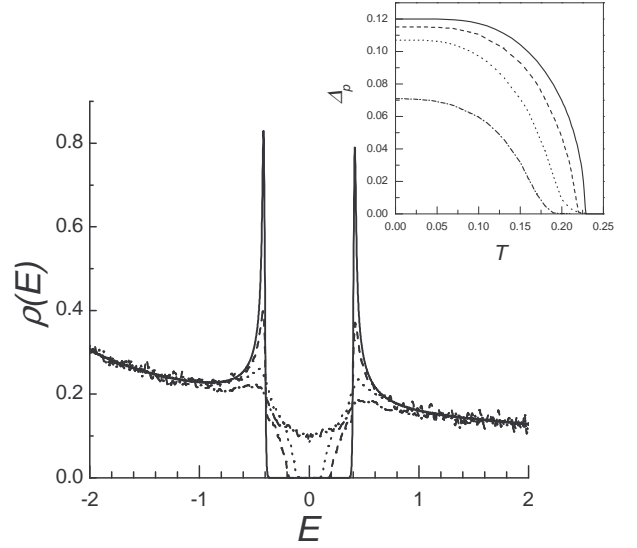}
\caption{Density of states as a function of energy of the
disordered $p_x+ip_y$-wave pairing state for the clean limit
(solid line), $n_{\text{imp}}=2\%$, $U_0=2$ (dashed line),
$n_{\text{imp}}=5\%$, $U_0=2$ (dotted line), $n_{\text{imp}}=5\%$,
$U_0=10$ (dash-dotted line). Also shown in the inset is the
temperature dependence of the impurity-averaged order parameter
$\Delta_p(T)$.}%
\label{dos}
\end{figure}

Here we employ a mean-field Bogoliubov-de Gennes (BdG) Hamiltonian
on a tight-binding triangular lattice with the nearest-neighbor
(NN) hopping integral $t$ and the NN bond pairing potential
$\Delta_{ij}$ resulting from the effective attractive interaction
$V$. The model Hamiltonian \cite{qhan} is expressed as
\begin{eqnarray}
H_{\text{eff}}&=&-\sum_{i,j,\sigma }t_{ij}c_{i\sigma }^{\dagger
}c_{j\sigma}+\sum_{i,\sigma }(\epsilon_i-\mu )c_{i\sigma }^{\dagger }c_{i\sigma } \nonumber\\%
&&+\sum_{\langle i,j\rangle} \left [\Delta _{ij}(c_{i\uparrow
}^{\dagger }c_{j\downarrow }^{\dagger}+c_{i\downarrow }^{\dagger
}c_{j\uparrow }^{\dagger})+\text{h.c.}\right ],
\end{eqnarray}%
where $t_{ij}=t$ is the NN hopping integral and in the reminder of
this paper, we choose $t<0$ according to the analysis on the band
calculation \cite{singh} and the energies will be measured in unit
of $|t|$. $\mu$ is the chemical potential. Here, we adopt the
binary alloy disorder model \cite{atkinson} where $\epsilon_i$ is
the $\delta$-function-like {\it scalar} scattering potential and
takes the value $U_0$ on certain lattice sites according to the
impurity concentration $n_{\text{imp}}$ and zero elsewhere. The
spin-triplet pairing potential $\Delta_{ij}$ is defined as $
\Delta_{ij}=\frac{V}{2}(\langle
c_{i\uparrow}c_{j\downarrow}\rangle+\langle
c_{i\downarrow}c_{j\uparrow}\rangle)$. Note that only the $d_z$
component of the spin-triplet pairing is considered here in view
of the experimental indication that the $\bm{d}$ vector is
parallel to $z$-axis according to the invariant behavior of the
Knight shift for the in-plane magnetic field \cite{waki}. In the
homogenous
case, the $p_x+ip_y$-wave pairing state is expressed as %
\begin{eqnarray}
\Delta_{p_x+ip_y}(\bm{k})&=& 2\Delta_p\left[
\sin(k_x)+\sin(k_x/2)\cos(\sqrt{3}k_y/2)+ \right. \nonumber \\%
&& \left. i\sqrt{3}\cos(k_x/2)\sin(\sqrt{3}k_y/2) \right]
\label{deltak}
\end{eqnarray}
where
$\Delta_p=\frac{1}{6N}\sum_{i,\delta}\Delta_{ii+\delta}e^{-i\theta(\delta)}$
with $i+\delta$ the six NN sites of $i$. By applying the
self-consistent mean-field approximation and performing the
Bogoliubov transformation, diagonalization of the Hamiltonian
$H_{\text{eff}}$ can be achieved by solving the following BdG
equations:
\begin{equation}
\sum_{j}\left(
\begin{array}{cc}
H_{ij} & \Delta_{i,j} \\
-\Delta _{i,j}^{*} & -H_{ij}^{*}
\end{array}%
\right) \left(
\begin{array}{c}
u_{j}^{n} \\
v_{j}^{n}%
\end{array}%
\right) =E_{n}\left(
\begin{array}{c}
u_{j}^{n} \\
v_{j}^{n}
\end{array}%
\right)  \label{BdG}
\end{equation}%
where $u^{n},v^{n}$ are the Bogoliubov quasiparticle amplitudes \
with corresponding eigenvalue $E_{n}$.
$H_{ij}=-t_{ij}+\delta_{i,j}(\epsilon_i-\mu)$. $\Delta_{ij}$ is
calculated according to: $ \Delta_{ij}=\frac{V}{4}\sum_{n}(u_{i}^n
v_{j}^{n*}+ u_{j}^n v_{i}^{n*})\tanh(\frac{E_n}{2k_B T})$.
Throughout this work, we set $V=2.3$ and $\mu=1.0$ which gives
rise to $\Delta_p=0.12$ and the electron number per site is $1.39$
in the absence of disorder. Due to the vanishingly small
anisotropic factor of the $p_x+ip_y$-wave pairing \cite{comment1},
there is one $s$-wave-like full gap opened at approximately
$\Delta_{\text{Gap}}\approx 0.4$ (see Fig.~\ref{dos} for the clean
limit).

Once the self-consistent quasiparticle spectrum is obtained, the
nuclear spin relaxation rate is calculated according to
\cite{leadon},
\begin{eqnarray}
R(i,j)&=&\text{Im}\chi_{-,+}(i,j,i\Omega_n\rightarrow\Omega+i0^{+})%
/(\Omega/T)|_{\Omega\rightarrow 0} \nonumber \\
&=&\pi\int\int[\rho_{11}^{ij}(E)\rho_{22}^{ij}(-E^\prime)
-\rho_{12}^{ij}(E)\rho_{21}^{ij}(-E^\prime)] \nonumber \\
&&\times f(E)[1-f(E^\prime)]\delta(E-E^\prime)dEdE^\prime, \\
R(i,i)&=&-\pi
T\int_{-\infty}^{\infty}\rho_{11}^{ii}(E)\rho_{22}^{ii}(-E)
f^\prime(E)dE \label{Rii}
\end{eqnarray}
where $\rho_{\alpha\beta}^{ij}(E)$ is expressed as%
\begin{equation}
\left(
\begin{array}{cc}
\rho_{11}^{ij}(E) & \rho_{12}^{ij}(E) \\
\rho_{21}^{ij}(E) & \rho_{22}^{ij}(E)
\end{array}
\right)=\sum_n
\left(
\begin{array}{cc}
u_i^n u_j^{n*} & u_i^n v_j^{n*} \\
v_i^n u_j^{n*} & v_i^n v_j^{n*}
\end{array}
\right)\delta(E-E_n). \nonumber
\end{equation}
For the unconventional pairing, the contributions from the
off-diagonal elements $\rho_{12}$ and $\rho_{21}$ are absent in
Eq.~(\ref{Rii}) due to the pairing symmetry which forbids the
on-site pairing amplitude \cite{matsumoto}. The local density of
states (DOS) is calculated according to
\begin{equation}
\rho(i,E)=2\rho_{11}^{ii}(E)=2\rho_{22}^{ii}(-E).
\end{equation}

To numerically investigate the disorder effect on the electronic
structure and accordingly the nuclear spin relaxation rate in the
2D system, for certain impurity content $n_\text{imp}$, typical
DOS and $1/T_1$ are obtained by averaging over $20$ impurity
configurations with the size of the supercell $20\times 20$ and
$100$ wave vectors in the supercell Brillouin zone\cite{atkinson}.
For each impurity configuration, self-consistent bond pairing
potential $\Delta_{ij}$ is obtained with the maximum relative
error between two consecutive iteration steps is less than
$10^{-3}$. In this work, assuming that $1/T_1(i)$ is determined by
the on-site $R(i,i)$ with ignoring the minor contributions from
neighboring sites $R(i,i+\delta)$, we have the impurity-averaged
$1/T_1=\overline{T_1^{-1}(i)}$ and $\rho(E)=\overline{\rho(i,E)}$,
where $\overline{(\ldots)}$ denotes averaging over space and
impurity configurations.

In Fig.~\ref{dos}, we illustrate the dependence of the
disorder-averaged DOS on the impurity content $n_{imp}$ and the
scattering strength $U_0$. The set of curves  display clearly (1)
the shrinking (and even vanishing) of the energy gap (2) the
smearing and decreasing of the coherence peak as $n_{\text{imp}}$
and $U_0$ increase. As $n_{\text{imp}}$ increases from 0 to $2\%$
and $5\%$ for the fixed $U_0=2$, the gap is filled from the gap
edge, resulting in a reduced effective gap. And when $U_0=10$ for
$n_{\text{imp}}=5\%$, strictly speaking the energy gap is closed
with finite $\rho(E)$ as $E\rightarrow 0$ (although the DOS hump
around $\Delta_{\text{Gap}}$ can still be identified) and the
residual DOS at the Fermi level is as large as $60\%$ of the
normal state value. The large residual DOS in the superconducting
state is consistent with the experimental study of specific
heat\cite{ueland}, indicating the importance of inhomogeneity in
this material. In the inset of Fig.~\ref{dos}, we give
$\Delta_p(T)$ in the clean limit, $n_{imp}=2\%$, $U_0=2$,
$n_{imp}=5\%$, $U_0=2$ and $n_{imp}=5\%$, $U_0=10$, showing that
both the order parameter and the transition temperature are
reduced and the decrease of $\Delta_p$ as $T  \rightarrow
T_{\text{c}}$ is also not so sharp in the disordered cases as in
the dilute limit. These behaviors are in consistence with the pair
breaking effect of non-magnetic impurities in unconventional
superconductors.

The  evolution of the density of states originates from the
impurities, and depends on their content as well as strength.
 For the weak scalar scatterer with $U_0=2$
(corresponds to $c_s\simeq 0.5$ according to
Ref.~\onlinecite{qhwang1}), there are two peaks of the local DOS
around an isolated impurity near the gap edge with energies
$\omega_b/\Delta_{\text{Gap}}\simeq\pm 0.89$ highlighting the
presence of impurity bound state in the unconventional
$p_x+ip_y$-wave superconductor\cite{matsumoto,qhwang1} in
remarkable contrast to the conventional $s$-wave superconductors
although both are fully gapped. And for strong potential such as
$U_0=10$ (corresponds to $c_s\simeq 2.5$), we have
$\omega_b/\Delta_{\text{Gap}}\simeq\pm 0.37$. For finite impurity
density,
 to look into the closure of
the energy gap induced by the impurities, we study the
impurity-averaged self-energy of the quasiparticle, which is
determined by the self-consistent
equations \cite{pethick,hirschfeld1}: %
\begin{eqnarray}
G(\bm{k},i\omega_n)&=&[i\tilde{\omega}_n-\xi(\bm{k})\sigma_z
-\Delta_{p_x}\sigma_{x}-\Delta_{p_y}\sigma_{y}] \\
\Sigma(i\omega_n)&=&n_{\text{imp}}\frac{g(i\omega_n)}{U_0^{-2}-g^2(i\omega_n)}
\end{eqnarray}
where $i\tilde{\omega}_n=i\omega_n-\Sigma(i\omega_n)$.
$\Delta_{p_x}$ and $\Delta_{p_y}$ are real and imaginary parts of
Eq.~(\ref{deltak}), respectively. $\xi(\bm{k})$ is the normal
state quasiparticle energy and $g(i\omega_n)=\int
d^2{\bm{k}}/(2\pi)^2 G(\bm{k},i\omega_n)$. For the weak scatterers
$U_0^{-1}\gg g(i\omega_n)$, the scattering rate $\gamma$
[determined by $\Sigma(\omega\rightarrow 0)=-i\gamma$] is
$\gamma=\sqrt{(\pi N_0n_\text{imp}U_0^2)^2-\Delta_\text{Gap}^2}$,
where $N_0$ is the normal density of states per spin at the Fermi
level. Therefore, the impurity parameter $n_{\text{imp}}U_0^2$
must be larger than $\Delta_{\text{Gap}}(T)/\pi N_0$ to entirely
close the gap, i.e. $\gamma\neq 0$ is real. This effect of disorders for
the $p_x+ip_y$-wave pairing is significantly different from that
for the high-$T_\text{c}$ nodal $d_{x^2-y^2}$-wave pairing. In the
$d$-wave pairing, infinitesimal $n_{\text{imp}}U_0^2$ gives rise
to finite density of states at the Fermi level. In the strong
scattering limit, $U_0^{-1}\ll g(i\omega_n)$, we obtain
$\gamma=\sqrt{n_\text{imp}\Delta_\text{Gap}/\pi N_0}$, which is
the same as the result of $d$-wave superconductors with unitary
impurities \cite{hirschfeld2,balatsky}. The above discussion
qualitatively explain what we illustrate in Fig.~\ref{dos}.

\begin{figure}
\includegraphics[width=8 cm]{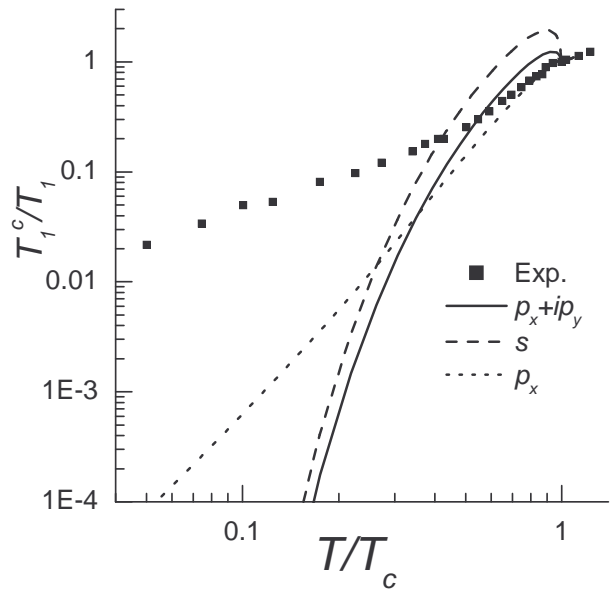}
\caption{Temperature dependence of $1/T_1$ normalized by its value
at $T_{\text{c}}$ for the $p_x+ip_y$-wave (solid line), $s$-wave
(dashed line), and $p_x$-wave (dotted line) superconductors. Solid
squares represent the experimental
data extracted from Fig.~2 of Ref.~\onlinecite{fujimoto}} %
\label{NMR-Pure}
\end{figure}%
The impurity effect on DOS is manifested by the variation of
NMR relaxation according to Eq.~(\ref{Rii}). First, we address the
temperature dependence of $1/T_1$ in the absence of disorder
(clean limit). When $n_{\text{imp}}=0$, the
$1/\sqrt{E^2-\Delta_{\text{Gap}}^2}$ divergence of the DOS
$\rho(E)$ at the gap edge $\Delta_{\text{Gap}}$ will lead to the
Hebel-Slichter coherence peak of $1/T_1$ just below $T_{\text{c}}$
as shown in Fig.~\ref{NMR-Pure}, although the jump of the peak is
much lower than that of the $s$-wave pairing because the coherence
factor changes from $1+\Delta_\text{Gap}^2/E^2$ in the $s$-wave
case to 1 in the $p_x+ip_y$-wave case. Furthermore, when $T\ll
T_{\text{c}}$ and $\Delta_\text{Gap}\gg T$, $1/T_1$ of both the
$s$-wave and $p_x+ip_y$-wave pairing states shows the behavior of
$e^{-\Delta_{\text{Gap}}/T}$ due to their fully gapped nature.
Also shown in Fig.~\ref{NMR-Pure} is the behavior of $1/T_1$ of
the gapless $p_x$-wave pairing state for comparison. As for the
nodal $p_x$-wave pairing [real part of Eq.~(\ref{deltak})], the
slower logarithmic divergent of $\rho(E)$ and the halved coherent
factor result in much suppressed coherence peak and in the low
temperature region $\rho(E)\propto E$ gives rise to the $T^3$
dependence of the nuclear spin relaxation rate as shown in the
figure.

\begin{figure}[t]
\includegraphics[width=8 cm]{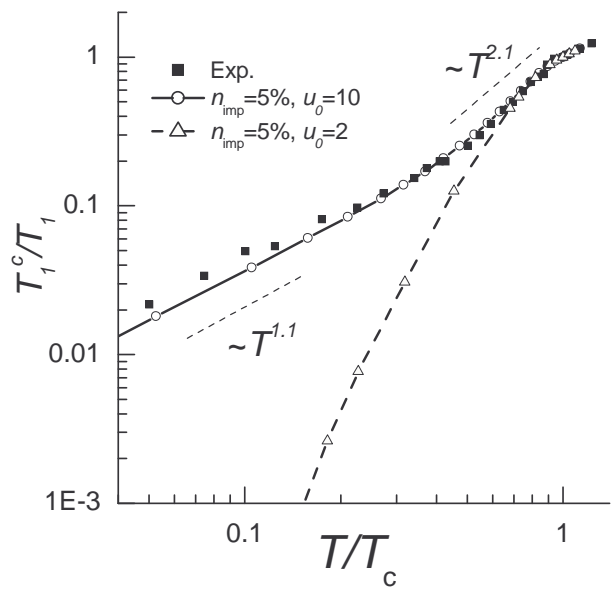}
\caption{Temperature dependence of $T_1^{\text{c}}/T_1$ for the
disordered $p_x+ip_y$-wave when $n_{\text{imp}}=5\%$, $U_0=10$
(solid line) and $n_{\text{imp}}=5\%$, $U_0=2$ (dashed line). The
Solid squares are the same as those of Fig.~\ref{NMR-Pure}} %
\label{NMR-Disorder}
\end{figure}%
Figure \ref{NMR-Disorder} shows  the nuclear spin relaxation rate
when the impurity scattering is present. Two typical results are
compared with the experimental data\cite{fujimoto}. As expected,
the coherence peak of $1/T_1$ disappears for both the weak and
strong scattering cases, in accord with the finding that small
concentration of impurities is able to smear the sharp divergence
of the DOS near the gap edge. For strong disorders with $U_0=10$,
our results of $1/T_1$ as a function of temperature indicate a
$T^{1.1}$ dependence at the low temperature region and a $T^{2.1}$
relation near $T_c$, being consistent with the results of an
approximate constant $\rho(E)$ near zero energy as shown in
Fig.~\ref{dos}.
 More importantly, these results are in good agreement with the experimental
observation of Ref.~\onlinecite{fujimoto}. On the other hand, the
study of weak disorders with $U_0=2$ exhibits that $1/T_1$ first
drops with a $T^n$ ($n\simeq 3$) law at the vicinity of $T_c$ and
then exponentially similar to its behavior in the clean limit
showing a {\it downward} curvature. The occurrence of the
exponential behavior is due to the opening of the energy gap at
certain temperature $T^*$ for the weak scatterers. And $T^*$ is
governed by the solution of $\Delta_\text{Gap}(T^*)=\pi N_0
n_\text{imp}U_0^2$, which results in $T^*/T_c\simeq 0.8$ for
$U_0=2$ and $n_\text{imp}=5\%$. According to this, one will have a
power-law dependence of $1/T_1$ down to lower temperature region
by simply increasing the impurity content $n_\text{imp}$.

In summary, we have for the first time elucidated the disorder effects on the
electronic structure and nuclear spin relaxation rate of the
$p_x+ip_y$-wave pairing state, which is  closely relevant
to the new superconductor Na$_{0.35}$CoO$_2$$\cdot y$H$_2$O.
The experimentally observed
temperature dependence of $1/T_1$ is explained satisfactorily.
It is also interesting to compare
the present results with those for another layered superconductor ,
Sr$_2$RuO$_4$\cite{miyake,ishida} with a possible
$p_x+ip_y$-wave pairing symmetry.
 We found that the gap in Na$_{0.35}$CoO$_2$$\cdot y$H$_2$O is highly
isotropic\cite{comment1} in the triangular lattice,
while the gap in Sr$_2$RuO$_4$ with the proposed $p_x+ip_y$-wave pairing
is  strongly anisotropic\cite{miyake}. Therefore, we
predict that future $1/T_1$ measurements on high-quality samples of
Na$_{0.35}$CoO$_2$$\cdot y$H$_2$O may observe both the coherence
peak and the exponential dependence on temperature, in contrast to
the $T^3$ \cite{ishida} behavior of $1/T_1$ in the significantly
anisotropic Sr$_2$RuO$_4$ even if both are in the hypothetical
$p_x+ip_y$-wave pairing states.

On the other hand, as far as the DOS-related physical quantities,
such as $1/T_1$, are concerned, there should be no qualitative
differences between the $p_x+ip_y$-wave pairing and the
$d+id^\prime$-wave pairing state in the sense that they are both
fully-gapped and have similar response to
impurities\cite{qhwang1}. Therefore, more definitive experimental
measurements of the Knight shift are demanded to determine the
symmetry of the spin part of the Cooper pair wave function.
Moreover, the muon-spin-relaxation measurement \cite{luke}, which
is sensitive to the time-reversal-symmetry breaking effect in the
$p_x+ip_y$- and $d+id^\prime$-wave pairing states, and the
phase-sensitive Josephson-tunnelling related experiments
\cite{tsuei} can give more decisive evidences to distinguish the
gapped pairing states from the nodal ones, such as the $p_x$ and
$d_{x^2-y^2}$ waves. The scanning tunnelling microscopy
experiments \cite{pan} are also able to shed light on the pairing
symmetry by examining the energy and spatial pattern of the
impurity \cite{qhwang1} and vortex \cite{jxzhu2,qhan} states.

We thank  Y. Chen, J.-X. Li, X.-G. Li, and   Q.-H. Wang for  helpful discussions.
We appreciate G.-q. Zheng for his sending us the preprint of Ref.[10].
 The work was
supported by the RGC grant of Hong Kong under Grant Nos.
HKU7092/01P and HKU7075/03P, the NSFC under Grant No. 10204019,
and the 973-project of the Ministry of Science and
Technology of China under Grant No. G1999064602.


\begin{thebibliography}{99}
\bibitem{comment1}
We find that for $\bm{k}$ on the Fermi surface, the gap anisotropy
$|\Delta_{p_x+ip_y}(\bm{k})|^\text{max}/|\Delta_{p_x+ip_y}(\bm{k})|^\text{min}$
varies from 1.08 (for $0.2$ electron dopping) to 1.02 (for $0.4$
electron dopping) for $t<0$.
\end{thebibliography}
\end{document}